# Nine Theorems of Inconsistency in GRT with Resolutions via *Isogravitation*

Ruggero Maria Santilli
Institute for Basic Research, P.O. Box 1577, Palm Harbor, FL 34682
e-mails ibr@verizon.net, ibr@gte.net

This paper presents nine inconsistency theorems for general relativity theory (GRT), and shows that they ultimately originate from the use of Riemannian curvature and the abandonment of universal invariance (which is stronger than the customary covariance). These features cause GRT to be non-canonical at the classical level and non-unitary at the operator level, resulting under time evolution in catastrophic structural problems, such as lack of invariant basic measurement units, loss of stable numerical predictions, absence of reliable observables, *etc*. The nine inconsistency theorems are re-inspected via *isotopic* methods and the related new theory of gravitation known as *isogravitation*. This new theory offers swift and simple resolutions for all the inconsistencies identified in GRT.

## 1. Introduction

It is well known that electroweak theories have an outstanding scientific consistency (see, *e.g.*, Refs. [1]), but, despite attempts dating back to Einstein, the achievement of a grand unification with the inclusion of gravity as represented by general relativity theory (GRT) [2] has remained elusive.

Previously published works [3] have pointed out a number of axiomatic inconsistencies of grand unifications in the representation of matter, as well as of antimatter, whenever gravity is represented via curvature in a Riemannian space. These include:
**1)** The admission by electroweak interactions of the fundamental Poincaré symmetry, compared to the absence of a symmetry for any Riemannian treatment of gravitation, in favor of the well known covariance;
**2)** The essentially flat, thus canonical, structure of electroweak interactions, compared to the curved, thus non-canonical, structure of Riemannian gravitation, with resulting non-unitary character of quantum gravity and related well-known problems of consistency;
**3)** The admission by electroweak interactions of negative-energy solutions for antimatter, as compared to the strict absence of negative energies for any Riemannian treatment of gravitation.

An axiomatically consistent grand unification was then attempted in Refs. [3] via the iso-Minkowskian representation of gravity [4] because: **i)** iso-Minkowskian gravity admits a symmetry for matter that is isomorphic to the Poincaré symmetry, thus resolving inconsistency 1); **ii)** iso-Minkowskian gravity replaces the Riemannian curvature with a covering notion compatible with the flatness of electroweak theories, thus resolving inconsistency 2); and **iii)** inconsistency 3) is resolved via the isodual theories of antimatter [5], including the isodual iso-Minkowskian geometry [5g] that permits negative-energy solutions for the gravitational field of antimatter.

The present work presents a critical analysis of the axiomatic foundations of GRT via the study of nine theorems dealing with inconsistencies so serious as to be at times known as 'catastrophic'; that is, apparently requiring the abandonment of the representation of gravity via curvature in favor of broader views.

## 2. SRT Consistency and Limitations

Thanks to historical contributions by Lorentz, Poincaré, Einstein, Minkowski, Weyl, and others (see, *e.g.*, the historical accounts [2f, 2g]), special relativity theory (SRT) achieved a majestic axiomatic and physical consistency. After a century of studies, we can safely identify the origins of this consistency in the following crucial properties:
**1)** SRT is formulated in the Minkowski spacetime $M(x, \eta, R)$ with local spacetime coordinates, metric, line element and basic unit given respectively by

$$x = \{x^\mu\} = (r^k, t), \ k = 1, 2, 3, \mu = 1, 2, 3, 0, \ c_0 = 1 \quad (2.1a)$$

$$\eta = \mathrm{Diag.}(1, 1, 1, -1) \quad (2.1b)$$

$$(x - y)^2 = (x^\mu - y^\mu) \times \eta_{\mu\nu} \times (x^\nu - y^\nu) \quad (2.1c)$$

$$I = \mathrm{Diag.}(1, 1, 1, 1) \quad (2.1d)$$

over the field of real numbers $R$, where we identify the conventional associative multiplication with the symbol $\times$ in order to distinguish it from the numerous additional multiplications used in the studies herein considered [3-10];
**2)** All laws of SRT, beginning with the above line element, are *invariant* (rather than covariant) under the fundamental *Poincaré symmetry*

$$\mathcal{P}(3.1) = \mathcal{L}(3.1) \times \mathcal{T}(3.1) \quad (2.2)$$

where $\mathcal{L}(3.1)$ is the *Lorentz group* and $\mathcal{T}(3.1)$ is the *Abelian group of translations* in spacetime; and
**3)** The Poincaré transformations are *canonical* at the classical level and *unitary* at the operator level, with implications crucial for physical consistency, such as the invariance of the assumed basic units (as per the very definition of a canonical or unitary transformation),

$$\mathcal{P} \times [\text{Diag.}(1\ \text{cm},\ 1\ \text{cm},\ 1\ \text{cm},\ 1\ \text{sec})] \times \mathcal{P}^{\text{tr}} \equiv \text{Diag.}(1\ \text{cm},\ 1\ \text{cm},\ 1\ \text{cm},\ 1\ \text{sec}) \quad (2.3)$$

with the resulting fundamental property that *SRT admits basic units and numerical predictions that are invariant in time*. In fact, the quantities characterizing the dynamical equations are the *Casimir invariants* of the Poincaré symmetry.

As a result of the above features, special relativity has been and can be confidently applied to experimental measurements because the units selected by the experimenter do not change in time, and the numerical predictions of the theory can be tested at any desired time under the same conditions without fear of internal axiomatic inconsistencies.

Despite these historical results, it should be stressed that, as is the fate for all theories, *SRT has its own well-defined limitations*. To begin, even within the indicated conditions of its original conception, SRT may well result not to be uniquely applicable due to its apparent biggest limitation, the inability to admit an absolute reference frame associated to the *ether* as a universal medium needed for the characterization and propagation of electromagnetic waves.

In fact, not only do electromagnetic waves need the ether for their existent formulation, but also elementary particles, such as the electron, are known to be mere oscillations of said universal medium. Rather than being forgotten just because vastly ignored, the issue of the privileged reference frame, and its relationship to the reference frames of our laboratories, is more open than ever, and may eventually force the use of an alternative formulation of SRT.

SRT is also inapplicable for the *classical treatment of antiparticles*, as shown in detail in Ref. [5g] and monograph [7f]. This is essentially due to the existence of only one quantization channel. Therefore, the quantization of a classical antiparticle characterized by special relativity (essentially that via the sole change of the sign of the charge) clearly leads to a quantum 'particle' with the wrong sign of the charge, and definitely not to the appropriate charge-conjugated state, resulting in endless inconsistencies.

At any rate, the insufficiency of SRT for the classical treatment of antimatter can be seen from the absence of any distinction between neutral particles and their antiparticles, a feature that propagates at the gravitational level, resulting in the current, virtually complete absence of quantitative studies as to whetehr distant (neutral) galaxies and quasars are made up of matter or of antimatter.

In fact, the achievement of the correct antiparticle at the quantum level has required the construction of the new *isodual mathematics* as an anti-isomorphic image of conventional mathematics, including its own *isodual quantization* and, inevitably, the construction of the new *isodual SRT* (for brevity, see Ref. [7d] and quoted literature). In this case, the isodual characterization of a classical antiparticle does indeed lead, under the isodual (rather than conventional) quantization, to the correct antiparticle as a charge conjugated state.

Special relativity has also been shown to be *inapplicable* (rather than violated) for the treatment of both, particles and antiparticles, such as hadrons, represented as they are in physical reality: extended, generally non-spherical and deformable (such as protons or antiprotons), particularly when interacting at very short distances. In fact, these conditions imply the mutual penetration of the wave-packets and/or the hyper-dense media constituting the particles, resulting in non-local integro-differential interactions that cannot be entirely reduced to potential interactions among point-like constituents

The historical inability of SRT to represent irreversible processes should also be recalled, and identified in the reversibility of the mathematical methods used by SRT, the reversibility in time of its basic axioms being a mere consequence. An additional field of inapplicability of SRT is that for all biological entities, since the former can only represent perfectly rigid and perfectly reversible, thus eternal structures, while biological entities are notoriously deformable and irreversible, having a finite life.

Mathematical studies of these aspects can be found in Refs. [6], while comprehensive treatments appear in Refs. [7] (Ref. [7e] in particular). For independent works, see Refs. [8-10]).

It should be stressed that the above issues are not of purely academic interest, because they have a direct societal relevance in view of the increasingly cataclysmic climatic events facing mankind, with resulting need of new clean energies and fuels.

In fact, it is well known that, beginning from the combustion of carbon dating back to prehistoric ages, *all energy-releasing processes are irreversible*. Hence, the continued restriction of research on manifestly irreversible processes to verify a manifestly reversible theory, such as SRT, may jeopardize the orderly search for new clean energies and fuels, thus mandating the laborious search for a suitable irreversible covering of SRT [7e].

Note that the use of the words 'violation of special relativity' here would be inappropriate because SRT was specifically conceived for *point-like particles (and not antiparticles) moving in vacuum solely under retarded action-at-a-distance interactions* [2f]. As a matter of fact, antiparticles were still unknown at the time of the conception and construction of SRT. Similarly, states of deep mutual penetrations of extended hadrons, as occurring in the core of neutron stars or black holes, where simply unthinkable at the inception of special relativity.

## 3. GRT Inconsistencies due to Lack of Sources

Despite its widespread popular support, *GRT* has without doubt been, in contrast to SRT, the most controversial theory of the 20-th century. This Section and the next review some of the major mathematical, theoretical, and experimental inconsistencies of GRT, all published in the refereed technical literature, yet generally ignored by scientists in the field.

There exist subtle distinctions between 'Einstein's Gravitation', 'Riemannian formulation of gravity' and 'GRT' as it is used. For our needs, we here define *Einstein's gravitation* as the reduction of exterior gravitation in vacuum to pure geometry; namely, gravitation is solely represented via curvature in a Riemannian space $\mathcal{R}(x, g, R)$ with spacetime coordinates (2.1a) and a no-where-singular, real-valued, and symmetric metric $g(x)$ over the reals $R$, with field equations [2b,2c]

$$G_{\mu\nu} = R_{\mu\nu} - g_{\mu\nu} \times R / 2 = 0 \quad (3.1)$$

The right hand side is zero: as a central condition for Einstein's gravitation, *for a body with null* total electromagnetic field (i.e null total charge and null magnetic moment) in vacuum, there are no sources for the exterior gravitational field.

For our needs, we define as GRT any description of gravity on a Riemannian space over the reals with Einstein-Hilbert field equations, with a source due to the presence of electric and magnetic fields:

$$G_{\mu\nu} = R_{\mu\nu} - g_{\mu\nu} \times R/2 = k t_{\mu\nu} \quad (3.2)$$

Here $k$ is a constant depending on the selected unit whose value is here irrelevant. For the scope of this paper it is sufficient to assume that the *Riemannian description of gravity* coincides with GRT according to the above definition.

In the following, we shall first study the inconsistencies of Einstein gravitation; that is, first the inconsistencies in the entire reduction of gravity to curvature without source, and then the inconsistency of GRT; that is, the inconsistencies caused by curvature itself, even in the presence of sources.

It should be stressed that a technical appraisal of the content of this paper can be reached only following the study of the axiomatic inconsistencies of grand unified theories of electroweak and gravitational interactions whenever gravity is represented with curvature on a Riemannian space, irrespective of whether with or without sources [3].

*THEOREM 1 [11a]: Einstein's gravitation and GRT at large are incompatible with the electromagnetic origin of mass established by quantum electrodynamics, and thus they are inconsistent with experimental evidence.*

**Proof.** Quantum electrodynamics has established that the mass of all elementary particles, whether charged or neutral, has a primary electromagnetic origin; that is, all masses have a first-order origin given by the volume integral of the 00-component of the energy-momentum tensor $t_{\mu\nu}$ of electromagnetic origin,

$$m = \int d^4x \times t_{00}^{\text{elm}} \quad (3.3a)$$

$$t_{\alpha\beta} = \frac{1}{4\pi}\left(F_\alpha^\mu F_{\mu\beta} + \frac{1}{4} g_{\alpha\beta} F_{\mu\nu} F^{\mu\nu}\right) \quad (3.3b)$$

where $t_{\alpha\beta}$ is the *electromagnetic tensor*, and $F_{\alpha\beta}$ is the *electromagnetic field* (see Ref. [11a] for explicit forms of the latter with retarded and advanced potentials).

Therefore, quantum electrodynamics requires the presence of a *first-order source tensor* in the *exterior field equations* in vacuum, as in Eqs. (3.2). Such a source tensor is by conception absent from Einstein's gravitation (3.1). Consequently, Einstein's gravitation is incompatible with quantum electrodynamics.

The incompatibility of GRT with quantum electrodynamics is established by the fact that the source tensor in Eqs. (3.2) is of *higher order in magnitude*, thus being ignorable in first approximation with respect to the gravitational field, while according to quantum electrodynamics, said source tensor is of first order, thus not being ignorable in first approximation.

The inconsistency of both Einstein's gravitation and GRT is finally established by the fact that, for the case when the total charge and magnetic moment of the body considered are null, Einstein's gravitation and GRT allow no source at all. By contrast, as illustrated in Ref. [11a], quantum electrodynamics requires a first-order source tensor even when the total charge and magnetic moments are null due to the charge structure of matter. **q.e.d.**

The first consequence of the above property can be expressed via the following:

*COROLLARY 1A [11a]: Einstein's reduction of gravitation in vacuum to pure curvature without source is incompatible with physical reality.*

A few comments are now in order. As is well known, the mass of the electron is entirely of electromagnetic origin, as described by Eq. (3.3), therefore requiring a first-order source tensor in vacuum as in Eqs. (3.2). Therefore, Einstein's gravitation for the case of the electron is inconsistent with Nature. Also, the electron has a point charge. Consequently, *the electron has no interior problem at all, in which case the gravitational and inertial masses coincide*,

$$m_{\text{electron}}^{\text{grav.}} \equiv m_{\text{electron}}^{\text{iner.}} \quad (3.4)$$

Next, Ref. [11a] proved Theorem 1 for the case of a neutral particle by showing that the $\pi^0$ meson also needs a first-order source tensor in the exterior gravitational problem in vacuum since its structure is composed of one charged particle and one charged antiparticle in highly dynamic conditions.

In particular, the said source tensor has such a large value to account for the entire *gravitational mass* of the particle [11a]

$$m_{\pi^0}^{\text{grav.}} = \int d^4x \times t_{00}^{\text{elm}} \quad (3.5)$$

For the case of the interior problem of the $\pi^0$, we have the additional presence of short-range weak and strong interactions, representable with a new tensor $\tau_{\mu\nu}$. We, therefore, have the following:

*COROLLARY 1B [11a]: In order to achieve compatibility with electromagnetic, weak and strong interactions, any gravitational theory must admit two source tensors, a traceless tensor for the representation of the electromagnetic origin of mass in the exterior gravitational problem, and a second tensor to represent the contribution to interior gravitation of the short range interactions according to the field equations*

$$G_{\mu\nu}^{\text{int.}} = R_{\mu\nu} - g_{\mu\nu} \times R/2 = k \times \left(t_{\mu\nu}^{\text{elm}} + \tau_{\mu\nu}^{\text{short range}}\right). \quad (3.6)$$

A main difference between the two source tensors is that the electromagnetic tensor $t_{\mu\nu}^{\text{elm}}$ is notoriously traceless, while the second tensor $\tau_{\mu\nu}^{\text{short range}}$ is not. A more rigorous definition of these two tensors will be given shortly.

It should be indicated that, for a possible solution of Eqs. (3.6), various explicit forms of the electromagnetic fields, as well as of the short range fields originating the electromagnetic and short-range energy momentum tensors, are given in Ref. [11a].

Since both source tensors are positive-definite, Ref. [11a] concluded that the interior gravitational problem characterizes the *inertial mass* according to the expression

$$m^{\text{iner}} = \int d^4x \times \left(t_{00}^{\text{elm}} + \tau_{00}^{\text{short range}}\right) \quad (3.7)$$

with the resulting general law

$$m^{\text{inert.}} \geq m^{\text{grav.}} \quad (3.8)$$

where the equality solely applies for the electron.

Finally, Ref. [11a] proved Theorem 1 for the exterior gravitational problem of a neutral massive body, such as a star, by showing that the situation is essentially the same as that for the $\pi^0$. The sole difference is that the electromagnetic field requires the sum of the contributions from *all* elementary constituents of the star,

$$m_{\text{star}}^{\text{grav.}} = \sum_{p=1,2,...} \int d^4x \times t_{p00}^{\text{elem.}} \quad (3.9)$$

In this case, Ref. [11a] provided methods for the approximate evaluation of the sum that resulted to be of first-order also for stars with null total charge.

When studying a charged body, there is no need to alter Eqs. (3.6), since that particular contribution is automatically contained in the indicated field equations.

Once the incompatibility of GRT at large with quantum electrodynamics has been established, the interested reader can easily prove the incompatibility of GRT with quantum field theory and quantum chromodynamics, as implicitly contained in Corollary 1B.

An important property, apparently first reached in Ref. [11a] in 1974, is the following:

*COROLLARY 1C [11a]: The exterior gravitational field of a mass originates entirely from the total energy-momentum tensor (3.3b) of the electromagnetic field of all elementary constituents of said mass.*

In different terms, a reason for the failure to achieve a 'unification' of gravitational and electromagnetic interactions, initiated by Einstein himself, is that the said interactions can be 'identified' with each other, and, as such, they cannot be unified. In fact, in all unifications attempted until now, the gravitational and electromagnetic fields preserve their identity, and the unification is attempted via geometric and other means resulting in redundancies that eventually cause inconsistencies.

Note that conventional electromagnetism is represented with the tensor $F_{\mu\nu}$ and related Maxwell's equations. When electromagnetism is identified with exterior gravitation, it is represented with the energy-momentum tensor $t_{\mu\nu}$ and related Eqs. (3.6).

In this way, *gravitation results as a mere additional manifestation of electromagnetism*. The important point is that, besides the transition from the field tensor $F_{\mu\nu}$ to the energy-momentum tensor $T_{\mu\nu}$, there is no need to introduce a new interaction to represent gravity.

Note finally the irreconcilable alternatives emerging from the studies herein considered:
ALTERNATIVE I. Einstein's gravitation is assumed as being correct, in which case quantum electrodynamics must be revised in such a way as to avoid the electromagnetic origin of mass; or
ALTERNATIVE II: Quantum electrodynamics is assumed as being correct, in which case Einstein's gravitation must be irreconcilably abandoned in favor of a more adequate theory.

By remembering that quantum electrodynamics is one of the most solid and experimentally verified theories in scientific history, it is evident that the rather widespread assumption of Einstein's gravitation as having final and universal character is non-scientific.

THEOREM 2 [11b,7d]: *Einstein's gravitation (3.1) is incompatible with the Freud identity of the Riemannian geometry, thus being inconsistent on geometric grounds.*
**Proof**. The Freud identity [11b] can be written

$$R_\beta^\alpha - \frac{1}{2} \times \delta_\beta^\alpha \times R - \frac{1}{2} \times \delta_\beta^\alpha \times \Theta \\ = U_\beta^\alpha + \partial V_\beta^{\alpha\rho} / \partial x^\rho = k \times (t_\beta^\alpha + \tau_\beta^\alpha) \quad (3.10)$$

where

$$\Theta = g^{\alpha\beta} g^{\gamma\delta} \left( \Gamma_{\rho\alpha\beta} \Gamma_{\gamma\beta}^\rho - \Gamma_{\rho\alpha\beta} \Gamma_{\gamma\delta}^\rho \right) \quad (3.11a)$$

$$U_\beta^\alpha = -\frac{1}{2} \frac{\partial \Theta}{\partial g_\rho^{\rho\alpha}} g^{\gamma\beta} \uparrow_\gamma \quad (3.11b)$$

$$V_\beta^{\alpha\rho} = \frac{1}{2} \Big[ g^{\gamma\delta} \left( \delta_\beta^\alpha \Gamma_{\gamma\delta}^\rho - \delta_\beta^\rho \Gamma_{\gamma\delta}^\alpha \right) + \\ + \left( \delta_\beta^\rho g^{\alpha\gamma} - \delta_\beta^\alpha g^{\rho\gamma} \right) \Gamma_{\gamma\delta}^\delta + g^{\rho\gamma} \Gamma_{\beta\gamma}^\alpha - g^{\alpha\gamma} \Gamma_{\beta\gamma}^\rho \Big] \quad (3.11c)$$

Therefore, the Freud identity requires two first order source tensors for the exterior gravitational problems in vacuum, as in Eqs. (3.6) of Ref. [11a]. These terms are absent in Einstein's gravitation (3.1) that, consequently, violates the Freud identity of the Riemannian geometry. **q.e.d.**

By noting that trace terms can be transferred from one tensor to the other in the r.h.s. of Eqs. (3.10), it is easy to prove the following:

*COROLLARY.2A [7d]: Except for possible factorization of common terms, the $t$- and $\tau$-tensors of Theorem 2 coincide, respectively, with the electromagnetic and short range tensors of Corollary 1B.*

A few historical comments regarding the Freud identity are in order. It has been popularly believed throughout the 20-th century that the Riemannian geometry possesses only *four identities* (see, *e.g.*, Ref. [2h]). In reality, Freud [11b] identified in 1939 a *fifth identity* that, unfortunately, was not aligned with Einstein's doctrines and, as such, the identity was virtually ignored in the entire literature on gravitation of the 20-th century.

However, as repeatedly illustrated by scientific history, structural problems simply do not disappear with their suppression, and actually grow in time. In fact, the Freud identity did not escape Pauli, who quoted it in a footnote of his celebrated book of 1958 [2g]. The present author became aware of the Freud identity via an accurate reading of Pauli's book (including its important footnotes), and assumed the Freud identity as the geometric foundation of the gravitational studies presented in Ref. [7d].

Subsequently, in his capacity as Editor in Chief of **Algebras, Groups and Geometries**, the present author requested the mathematician Hanno Rund, a known authority in Riemannian geometry [2i], to inspect the Freud identity for the purpose of ascertaining whether the said identity was indeed a new identity. Rund kindly accepted Santilli's invitation and released paper [11c] of 1991 (the last paper prior to his departure) in which Rund indeed confirmed the character of Eqs. (3.10) as a genuine, independent, fifth identity of the Riemannian geometry.

The Freud identity was also rediscovered by Yilmaz (see Ref. [11d] and papers quoted therein), who used the identity for his own broadening of Einstein's gravitation via an external *stress-energy tensor* that is essentially equivalent to the source tensor with non-null trace of Ref. [11a], Eqs. (3.6).

Despite these efforts, the presentation of the Freud identity to various meetings and several personal mailings to colleagues in gravitation, the Freud identity continues to remain generally ignored to this day, with very rare exceptions (Contact by colleagues concerning additional studies on the Freud identify not quoted herein would be gratefully appreciated.)

Theorems 1 and 2 complete the presentation on the catastrophic inconsistencies of Einstein's gravitation due to the lack of a first-order source in the exterior gravitational problem in vacuum. Theorems 1 and 2 by no means exhaust all inconsistencies of Einstein's gravitation, and numerous additional inconsistencies do indeed exist. For instance, Yilmaz [11d] has proved that Einstein's gravitation explains the 43" of the precession of Mercury, but cannot explain the basic Newtonian contribution. This result can also be seen from Ref. [11a] because the lack of source implies the impossibility of importing into the theory the basic Newtonian potential. Under these conditions, the representation of the Newtonian contribution is reduced to a religious belief, rather than a serious scientific statement.

For these and numerous additional inconsistencies of GRT we refer the reader to Yilmaz [11d], Wilhelm [11e-11g], Santilli [11h], Alfvén [11i-11j], Fock [11k], Nordensen [11l], and large literature quoted therein.

## 4. GRT Inconsistencies due to Curvature

We now pass to the study of the structural inconsistencies of GRT caused by the very use of the Riemannian *curvature*, irrespective of the selected field equations, including those fully compatible with the Freud identity.

*THEOREM 3 [11m]: Gravitational theories on a Riemannian space over a field of real numbers do not possess time invariant basic units and numerical predictions, thus having serious mathematical and physical inconsistencies.*
**Proof**. The map from Minkowski to Riemannian spaces is known to be *non-canonical*,

$$\eta = \text{Diag.}(1,\ 1,\ 1,\ -1) \rightarrow g(x) = U(x) \times \eta \times U(x)^\dagger \quad (4.1a)$$

$$U(x) \times U(x)^\dagger \neq I \quad (4.1b)$$

Thus, the time evolution of Riemannian theories is necessarily non-canonical, with resulting lack of invariance of the basic units of the theory in time, such as

$$I_{t=0} = \text{Diag.}(1\ \text{cm},\ 1\ \text{cm},\ 1\ \text{cm},\ 1\ \text{sec}) \rightarrow$$
$$I'_{t>0} = U_t \times I \times U_t^\dagger \neq I_{t=0} \quad (4.2)$$

The lack of invariance in time of numerical predictions then follows from the known 'covariance', that is, lack of time invariance of the line element. **q.e.d.**

As an illustration, suppose that an experimentalist assumes at the initial time $t = 0$ the units 1 cm and 1 sec. Then, all Riemannian formulations of gravitation, including Einstein's gravitation, predict that at the later time $t > 0$ said units have a different numerical value.

Similarly, suppose that a Riemannian theory predicts a numerical value at the initial time $t = 0$, such as the 43" for the precession of the perihelion of Mercury. One can prove that the same prediction at a later time $t = 0$ is numerically different precisely in view of the 'covariance', rather than invariance as intended in special relativity, thus preventing a serious application of the theory to physical reality. We therefore have the following:

*COROLLARY 3A [11m]: Riemannian theories of gravitation in general, and Einstein's gravitation in particular, can at best describe physical reality at a fixed value of time, without a consistent dynamic evolution.*

Interested readers can independently prove the latter occurrence from the *lack of existence of a Hamiltonian in Einstein's gravitation*. It is known in analytic mechanics (see, *e.g.*, Refs. [2l, 7b]) that Lagrangian theories not admitting an equivalent Hamiltonian counterpart, as is the case for Einstein's gravitation, are inconsistent under time evolution, unless there are suitable subsidiary constraints that are absent from GRT.

It should be indicated that the inconsistencies are much deeper than that indicated above. For consistency, the Riemannian geometry must be defined on the field of numbers $R(n,+,\times)$ that, in turn, is fundamentally dependent on the basic unit $I$. But the Riemannian geometry does not leave time invariant the basic unit $I$ due to its non-canonical character. The loss in time of the basic unit $I$ then implies the consequential loss in time of the base field $R$, with consequential catastrophic collapse of the entire geometry [11m].

In conclusion, not only is Einstein's reduction of gravity to pure curvature inconsistent with Nature because of the lack of sources, but also the ultimate origin of the inconsistencies rests in the curvature itself when assumed for the representation of gravity, due to its inherent non-canonical character at the classical level with resulting non-unitary structure at the operator level.

Serious mathematical and physical inconsistencies are then unavoidable under these premises, thus establishing the impossibility of any credible use of GRT, for instance, as an argument against the test on antigravity predicted for antimatter in the field of matter [5], as well as establishing the need for a profound revision of our current views on gravitation.

*THEOREM 4: Observations do not verify Einstein's gravitation uniquely.*

**Proof:** All claimed 'experimental verifications' of Einstein's gravitation are based on the PPN 'expansion' (or linearization) of the field equations (such as the post-Newtonian approximation), that, as such, is not unique. In fact, Eqs. (3.1) admit a variety of inequivalent expansions, depending on the selected parameter, the selected expansion and the selected truncation. It is then easy to show that the selection of an expansion of the same equations (3.1) but different from the PPN approximation leads to dramatic departures from experimental values. **q.e.d**.

*THEOREM 5*: GRT is incompatible with experimental evidence because it does not represent the bending of light in a consistent, unique and invariant way.}

**Proof:** Light carries energy, thus being subjected to a bending due to the conventional Newtonian gravitational attraction, while Einstein's gravitation predicts that the bending of light is due to curvature (see, *e.g.*, Ref. [2h], Section 40.3). In turn, the absence of the Newtonian contribution causes other inconsistencies, such as the inability to represent the free fall where curvature does not exist (Theorem 6 below). Assuming that consistency is achieved with as yet unknown manipulations, the representation of the bending of light is not unique, because it is based on a nonunique PPN approximation having different parameters for different expansions. Finally, assuming that consistency and uniqueness are somewhat achieved, the representation is not invariant in time due to the noncanonical structure of GRT.

*THEOREM 6.* GRT is incompatible with experimental evidence because of the lack of consistent, unique and invariant representation of the free fall of test bodies along a straight radial line without curvature.

**Proof**: a consistent representation of the free fall of a mass along a straight radial line requires that the Newtonian attraction be represented the field equations necessarily without curvature, thus disproving the customary belief that said Newtonian attraction emerges at the level of the post-Newtonian approximation. **q.e.d.**

The absence from GRT at large, thus including Einstein's gravitation, of well defined contributions due to the Newtonian attraction and to the assumed curvature of spacetime, and the general elimination of the former in favor of the latter, cause other inconsistencies, such as the inability to represent the base Newtonian contribution in planetary motion as shown by Yilmaz [11d], and other inconsistencies [11e-11m].

A comparison between SRT and GRT is here in order. SRT can safely be claimed 'verified by experiments', because the said experiments verify numerical values uniquely and unambiguously predicted by SRT. By contrast, no such statement can be made for GRT, since the latter does not uniquely and unambiguously predict given numerical values, due, again, to the variety of possible expansions and linearization.

The origin of such a drastic difference is due to the fact that *the numerical predictions of SRT are rigorously controlled by the basic Poincaré invariance. By contrast, one of the several drawbacks of the 'covariance' of GRT is precisely the impossibility of predicting numerical values in a unique and unambiguous way, thus preventing serious claims of true 'experimental verifications' of GRT.*

By no means, the inconsistencies expressed by Theorems 3.1, 3.2, 4.1, 4.2 and 4.3 constitute all inconsistencies of GRT. In the author's opinion, additional deep inconsistencies are caused by the fact that *GRT does not possess a well defined Minkowskian limit*, while the admission of the Minkowski space as a tangent space is basically insufficient on dynamical grounds (trivially, because on said tangent space gravitation is absent).

As an illustration, we should recall the controversy on conservation laws that raged during the 20-th century [11]. Special relativity has rigidly defined total conservation laws because they are the Casimir invariants of the fundamental Poincaré symmetry. By contrast, there exist several definitions of total conservation laws in a Riemannian representation of gravity due to various ambiguities evidently caused by the absence of a symmetry in favor of covariance.

Moreover, none of the gravitational conservation laws yields the conservation laws of SRT in a clear and unambiguous way, precisely because of the lack of any limit of a Riemannian into the Minkowskian space. Under these conditions, the compatibility of GRT with SRT reduces to personal beliefs outside a rigorous scientific process. The above studies can be summarized with the following:

*THEOREM 7 [7d]: Gravitational theories on a Riemannian space cannot yield the conventional total conservation laws of SRT in a unique, unambiguous and invariant way due to lack of a unique, unambiguous and invariant Minkowskian limit.*

Another controversy that remained unresolved in the 20-th century (primarily because of lack of sufficient consideration by scholars in the field) is that, during its early stages, gravitation was divided into the *exterior and interior problems*. For instance, Schwartzchild wrote *two* articles on gravitation, one on the exterior and one on the interior problem [2d].

However, it soon became apparent that GRT was structurally unable to represent interior problems for numerous reasons, such as the impossibility of incorporating shape, density, local variations of the speed of light within physical media via the familiar law we study in high school $c = c_0 / n$ (which variation cannot be ignored classically), inability to represent interior contact interactions with a first-order Lagrangian, structural inability to represent interior non-conservation laws (such as the vortices in Jupiter's atmosphere with variable angular momenta), structural inability to represent entropy, its increase and other thermodynamic laws, *etc.* (see Ref. [7d] for brevity).

Consequently, Schwartzchild's solution for the *exterior* problem became part of history (evidently because aligned with GRT), while his *interior* solution has remained vastly ignored to this day (evidently because it is not aligned with GRT). In particular, the constituents of all astrophysical bodies have been abstracted as being point-like, an abstraction that is beyond the boundaries of science for classical treatments; all distinctions between exterior and interior problems have been ignored by the vast majority of the vast literature in the field; and gravitation has been tacitly reduced to one single problem.

Nevertheless, as indicated earlier, major structural problems grow in time when ignored, rather than disappearing. The lack of addressing the interior gravitational problem is causing major distortions in astrophysics, cosmology and other branches of science (see also next section). We have, therefore, the following important result:

*THEOREM 8 [7d]: GRT is incompatible with the experimental evidence on interior gravitational problems.*

By no means does the above analysis exhaust all inconsistencies of GRT, and numerous additional ones do indeed exist, such as that expressed by the following:

*THEOREM 9 [11m]: Operator images of Riemannian formulations of gravitation are inconsistent on mathematical and physical grounds.*

**Proof**. As established by Theorem 4.1, classical formulations of Riemannian gravitation are non-canonical. Consequently, all their operator counterparts must be non-unitary for evident reasons of compatibility. But non-unitary theories are known to be inconsistent on both mathematical and physical grounds [11m]. In fact, on mathematical grounds, non-unitary theories of quantum gravity (see, *e.g.*, Refs. [2j, 2k]) do not preserve in time the basic units, fields and spaces, while, on physical grounds, the said theories do not possess time invariant numerical predictions, do not possess time invariant Hermiticity (thus having no acceptable observables), and violate causality. **q.e.d.**

The reader should keep in mind the additional well known inconsistencies of quantum gravity, such as the historical incompatibility with quantum mechanics, the lack of a credible PCT theorem, *etc*. According to the ethics of science, all these inconsistencies should establish beyond a scientific doubt, or any otherwise credible doubt, the need for a profound revision of the gravitational views of the 20-th century.

## 5. Re-Inspetion of the Inconsistency Theorems via Isotopic Methods

In the author's view, the most serious inconsistencies in GRT are those of *experimental* character, such as the structural impossibility for the Riemannian geometry to permit unique and unambiguous numerical predictions due to the known large degrees of freedom in all PPN expansions; the necessary *absence* of curvature to represent consistently the free fall of bodies along a straight radial line; and the gravitational deflection of light measured until now being purely *Newtonian* in nature.

These inconsistencies are such to prevent serious attempts in salvaging GRT. For instance, if the deflection of the speed of light is re-interpreted as being solely due to curvature without any Newtonian contribution, then GRT admits other catastrophics inconsistencies, such as the inability to represent the Newtonian contribution of planetary motions pointed out by Yilmaz [11d], and other inconsistencies such as those identified by Wilhelm [11e-11g] and other researchers.

When the inconsistencies between GRT and experimental evidence are combined with the irreconcilable incompatibility of GRT with unified field theory and the catastrophics axiomatic inconsistencies due to lack of invariance [11m], time has indeed arrived for the scientific community to admit the need for fundamentally new vistas in our representation of gravitation, without which research is turned from its intended thrilling pursuit of 'new' knowledge to a sterile fanatic attachment to 'past' knowledge.

The nine inconsistency theorems identified in this paper define the axiomatic structure of the needed new gravitational theory, and quite rigidly so, as alternative is known to this author after decades of study. The only possible resolution of said inconsistency theorems requires that a new gravitational theory must satisfy the following requirements:

I. It must possess a single universal symmetry for all possible, interior and exterior gravitational models (to avoid the catastrophic inconsistencies caused by the conventional covariance and other reasons);

II. Said symmetry must be locally isomorphic to the Poincaré symmetry (to assure the true validity of conventional total conservation laws and other reasons); and

III. The new gravitational theory must admit a unique, unambiguous and invariant limit into SRT (as a basic compatibility condition of gravitation with SRT and other reasons).

To our best knowledge, the only new theory of gravitation capable of fulfiling the above conditions and bypassing all nine inconsistency theorems studied in this note (as well as resolve other inconsistencies omitted here for brevity) is that proposed by Santilli in Refs. [4] via the so-called *isotopic methods* (see Refs. [7c-7e] for comprehensive studies and references).

Alternatively, the latter methods provide an effective alternative study of the inconsistency theorems, such as those on total conservation laws (Theorem 7), interior gravitational problems (Theorem 8) and the inconsistencies of quantum gravity (Theorem 9) because the transparent and instantaneous solution provided by the isotopic methods confirms rather forcefully said inconsistency theorems.

The new theory of gravitation identified by the isotopic methods, and known as *isogravitation*, is based on the following simple main assumptions:

1) Factorization of any given (nonsingular) Riemannian metric $g(x)$ into a $4 \times 4$-dimensional matrix $\hat{T}(x) = \{\hat{T}_{\mu\nu}(x)\}$ and the conventional Minkowski metric $\eta$, Eq. (2.1b),

$$g(x) = \hat{T}(x) \times \eta \quad ; \qquad (5.1)$$

2) Assumption of the inverse of $\hat{T}(x)$ as the new basic unit of the theory in lieu of the conventional Minkowskian unit (2.1c),

$$\hat{I}(x) = \left[\hat{T}(x)\right]^{-1} \quad ; \qquad (5.2)$$

3) Reconstruction of the entire mathematical and physical setting of the *Minkowskian* (rather than the Riemannian) geometry in such a way as to admit $\hat{I}(x)$ (rather than $I$) as the new basic left and right unit at all levels.

Condition (3) is readily verified by lifting the conventional associative product $A \times B$ between two generic quantities $A, B$ into a new product under which $\hat{I}(x)$ is indeed the new right and left unit [12a,4a],

$$A \times B \rightarrow A \hat{\times} B = A \times \hat{T}(x) \times B \quad , \qquad (5.3a)$$

$$I \times A = A \times I \rightarrow \hat{I} \hat{\times} A = A \hat{\times} \hat{I} \equiv A \quad , \qquad (5.3b)$$

for all elements $A$ of the set considered.

The above liftings, called *isotopic* because they are axiom-preserving [12a], characterize a new mathematics today known as *Santilli isomathematics* [6], that includes new *isonumbers*, *isofields*, *isospaces*, *isofunctional analysis*, *isoalgebras*, *isogeometries*, *etc.* [7c-7d,10].

Since $\hat{I}(x)$ is positive-definite [from the assumed nonsingularity, the local Minkowskian character and factorization (5.1)], the resulting new spaces, first introduced in Ref. [4a] of 1983 and today known as [10] the *Minkowski-Santilli isospaces* $\hat{M}$, are locally isomorphic to the conventional space $M$.

Consequently, *the resulting Minkowski-Santilli isogeometry has no curvature, yet it admits all infinitely possible Riemannian line elements*.} Equivalent results can be reached by reformulating Riemannian line elements and related geometry (covariant derivative, Christoffel's symbols, *etc.*) with respect to the new unit $\hat{I}(x)$ (see memoir [4g] for geometric studies).

An important result is the achievement of the universal symmetry for all infinitely possible, locally Minkowskian, interior and exterior Riemannian line elements under the above reformulation, today known in the literature [10] as the Poincarè-Santilli isosymmetry $\hat{P}(3.1)$ (see: [4a,4b] for the first isotopies of the Lorentz symmetry at the classical and operator levels; [4c] for the first isotopies of the rotational symmetry; [4d] for the first isotopies of the SU(2)-spin symmetry; [4e] for the first isotopies of the Poincarè symmetry including the universal invariance of Riemannian line elements; and [4f] for the first isotopies of the spinorial covering of the Poincarè symmetry).

Since, again, $\hat{I}(x) > 0$, the new isosymmetry is locally isomorphic to the conventional one $\hat{P}(3.1) \approx P(3.1)$. In particular, *the generators of $\hat{P}(3.1)$ and $P(3.1)$ coincide, thus eliminating all controversies on total conservation laws ab initio*} (because, as recalled earlier, the rigorous formulation of conservation laws is that as generator of a symmetry, and certainly not of a covariance).

Moreover, the explicit form of the symmetry transformations (see [4] for brevity) is highly nonlinear, noncanonical and non-Lagrangian in conventional spacetime, yet the theory reconstructs linearity, canonicity and Lagrangian character in the Minkowski-Santilli isospace (for technical reasons interested readers have to study in the specialized literature).

Note the emergence of a unique, unambiguous and invariant limit from gravitational to relativistic settings given by

$$\text{Lim } \hat{I}(x) = I \qquad (5.4)$$

under which the entire Minkowskian formulations, including the conventional Poincarè symmetry, are recovered uniquely, identically and invariantly.

The resolution of the inconsistency theorems then follows not only from the elimination of curvature, but actually from a geometric unification of SRT and GRT via the axioms of the special, unification based on the embedding of gravitation where nobody looked for it, in the unit of relativistic theories} [4g].

An axiomatically consistent grand unification inclusive of gravitation is then another direct consequence, but only after working out a consistent *classical* theory of antimatter [3].

Another important consequence is the emergence of an axiomatically consistent operator theory of gravity that is reached, again, via the embedding of gravity in the unit of conventional relativistic quantum mechanics [12b], where consistency is guaranteed by the fact that the new theory is topologically equivalent to the conventional theory.

Intriguingly, Einstein-Hilbert field equations remain valid in the Minkowski-Santilli isogeometry, being merely reformulated via the new isomathematics, plus the addition of first-order sources for compatibility with relativistic treatments, Eqs. (3.6) [4g], and then we shall write in the isotopic form

$$\hat{G}^{\text{int}}_{\mu\nu} = \hat{R}_{\mu\nu} - \hat{T}(x)_{\mu\rho} \times \eta_{\rho\nu} \times \hat{R}/\hat{2} = \hat{k}\hat{\times}\left(\hat{t}^{\text{elm}}_{\mu\nu} + \hat{\tau}^{\text{short range}}_{\mu\eta}\right) \qquad (5.5)$$

where 'hat' indicates that the quantities are formulated on isospace over isofields.

The entire content of this paper and Refs. [3,4,12] can be restated by noting that the origin of the century-old controversies on GRT do not appear to be of physical nature, but rather of purely mathematical character because of originating from the treatment of gravitation with conventional mathematics. In fact, under the selection of a new mathematics more appropriate for the study of gravitation, all historical inconsistencies and controversies appear to be resolved while preserving Einstein-Hilbert equations in the reformulation (5.5) computed with respect to unit (5.2), in which case there is no curvature.

Physicists who are discouraged by new mathematics should be aware that the entire formalism of the new gravitation can be constructed very simply via the use of the following noncanonical/nonunitary transform

$$U \times U^{\dagger} = \hat{I}(x) > 0 \quad , \quad \hat{T}(x) = (U \times U^{\dagger})^{-1} > 0 \qquad (5.6)$$

where $\hat{I}(x)$ is the gravitational isounit (5.2), provided that it is applied to the *totality* of the formalism of the Minkowskian geometry, including unit $I$, numbers $c \in C$, products $A \times B$, functional analysis, metric spaces, Hilbert spaces, algebras, geometries, *etc.*, as illustrated below

$$I \to U \times I \times U^{\dagger} = \hat{I}(x) \quad , \qquad (5.7a)$$

$$c \to U \times c \times U^{\dagger} = \hat{m} = m \times \hat{I} \quad , \qquad (5.7b)$$

$$A \times B \to U(A \times B) \times U^{\dagger} = \hat{A}\hat{\times}\hat{B} \quad , \text{ etc.} \qquad (5.7c)$$

where for any $D = A, B$, *etc.*, $\hat{D} = U \times D \times U^{\dagger}$.

Invariance is easily proved by decomposing any additional noncanonical or nonunitary transforms in the isocanonical or isounitary form [6b],

$$W \times W^{\dagger} = \hat{I} \quad , \quad W = \hat{W} \times \hat{T}^{1/2} \quad , \qquad (5.8a)$$

$$W \times W^{\dagger} \neq I \quad , \quad \hat{W}\hat{\times}\hat{W}^{\dagger} = \hat{W}^{\dagger}\hat{\times}\hat{W} = \hat{I} \qquad (5.8b)$$

under which we have the following fundamental invariances

$$\hat{I} \to \hat{W}\hat{\times}\hat{I}\hat{\times}\hat{W} \equiv \hat{I} \quad , \qquad (5.9a)$$

$$\hat{A} \,\hat{\times}\, \hat{B} \rightarrow \hat{W} \,\hat{\times}\, (\hat{A} \,\hat{\times}\, \hat{B}) \,\hat{\times}\, \hat{W}^{\dagger} = \hat{A}' \,\hat{\times}\, \hat{B}' \ , \ etc. \quad (5.9b)$$

where $\hat{D}' = \hat{W} \,\hat{\times}\, D \,\hat{\times}\, \hat{W}^{\dagger}$, $D = A, B, ...$ and the invariance originates from the preservation of the isoproduct, $\hat{\times}' = \hat{\times}$ (since its change would imply passing from the assigned gravitational model characterized by $\hat{T}(x)$ to a *different}* gravitational model characterized by $\hat{T}'(x)$.)

In the hope of helping colleagues avoid writing papers that cannot stand the test of time, it should be stressed that *all the above results are crucially dependent on the "invariance of the basic gravitational unit"* Eqs. (5.2) and (5.9a). Isomathematics guarantees this invariance because, whether conventional or generalized, the unit is the basic invariant of any theory. This is why the unit (rather than any other quantity) was assumed for the only now-known consistent representation of gravitation.

The papers that cannot stand the test of time are those identifying a symmetry of Riemannian line elements, without the joint achievement of the invariance of the basic unit, in which case the nine theorems of catastrophic inconsistency due to noncanonical and nonunitary structure [11m] remain in full force and effect, despite the achievement of a symmetry.

To close with an intriguing historical note, Albert Einstein himself could be considered the initiator of the above isotopic formulation of gravity, because of his historical doubt on the *lack of completion of quantum mechanics}* [2m]. In fact, as illustrated in Eqs. (5.6)-(5.9), *the isotopic isotopic lifting of relativistic quantum mechanics, known as hadronic mechanics [3-12] and related gravitational content, is nothing but a completion via a nonunitary transform* (for the relationship of isotopies with the E-P-R argument, hidden variables, Bell's inequality, and related matters, see Ref. [12f]).

Needless to say, studies on the latter reformulation are only at their beginning, and so much remains to be done. It is hoped that some of the open problems can be treated in a follow-up paper.

## Acknowledgments

This paper grew out of numerous discussions at the biennial meetings *Physical Interpretations of Relativity Theories* organized at the Imperial College in London by the chapter of the British Society for the Philosophy of Sciences at the University of Sunderland, England. The author would like to express his deepest appreciation to the organizers of these meetings for their true scientific democracy, as well as to all its participants for openly expressing their views. Very special thanks for invaluable criticisms and comments are due to Professors A. Animalu, A. K. Aringazin, J. Dunning-Davies, P. Rowlands, H. Wilhelm and others. Additional thanks are due to Mrs. D. Zuckerman for an accurate linguistic control of the manuscript.